\documentclass[runningheads]{llncs}
\usepackage[utf8]{inputenc}
\usepackage{color}
\usepackage{subfig}
\usepackage{graphicx}
\usepackage{tabularx}
\usepackage{multirow}
\usepackage{booktabs}
\usepackage{amsmath}
\usepackage{amssymb}
\usepackage{xcolor}
\usepackage{arydshln}
\usepackage[]{algorithm2e}
\DeclareMathOperator*{\argmin}{arg\,min}

\newcolumntype{L}[1]{>{\raggedright\let\newline\\\arraybackslash\hspace{0pt}}m{#1}}
\newcolumntype{C}[1]{>{\centering\let\newline\\\arraybackslash\hspace{0pt}}m{#1}}
\newcolumntype{R}[1]{>{\raggedleft\let\newline\\\arraybackslash\hspace{0pt}}m{#1}}

\title{Extracting and Leveraging Nodule Features with Lung Inpainting for Local Feature Augmentation}
\titlerunning{Leveraging Nodule Features for Local Feature Augmentation}

\ifx\anonymize\undefined
\author{Sebastian G\"undel\inst{1,3}\and Arnaud A. A. Setio\inst{1}\and\\Sasa Grbic\inst{2}\and Andreas Maier\inst{3}\and Dorin Comaniciu\inst{2}}

\index{G\"undel, Sebastian}
\index{Setio, Arnaud}
\index{Grbic, Sasa}
\index{Maier, Andreas}
\index{Comaniciu, Dorin}

\authorrunning{S. G\"undel et al.}

\institute{Digital Technology and Innovation, Siemens Healthineers, Erlangen, Germany\and
Digital Technology and Innovation, Siemens Healthineers, Princeton, NJ, USA\and
Pattern Recognition Lab, Friedrich-Alexander-Universit\"at Erlangen, Germany
\email{sebastian.guendel@fau.de}
}

\else
\author{*}
\authorrunning{*}
\institute{*}
\fi

\begin{document}

\maketitle

\begin{abstract}
Chest X-ray (CXR) is the most common examination for fast detection of pulmonary abnormalities. Recently, automated algorithms have been developed to classify multiple diseases and abnormalities in CXR scans. However, because of the limited availability of scans containing nodules and the subtle properties of nodules in CXRs, state-of-the-art methods do not perform well on nodule classification. To create additional data for the training process, standard augmentation techniques are applied. However, the variance introduced by these methods are limited as the images are typically modified globally. In this paper, we propose a method for local feature augmentation by extracting local nodule features using a generative inpainting network. The network is applied to generate realistic, healthy tissue and structures in patches containing nodules. The nodules are entirely removed in the inpainted representation. The extraction of the nodule features is processed by subtraction of the inpainted patch from the nodule patch. With arbitrary displacement of the extracted nodules in the lung area across different CXR scans and further local modifications during training, we significantly increase the nodule classification performance and outperform state-of-the-art augmentation methods.
\end{abstract}

\keywords{Nodule classification \and Local feature augmentation \and Context encoder \and Chest X-ray}

\section{Introduction}

Lung cancer is one of the most frequent cancer worldwide. Combined with the high mortality rate, the efficiency of lung cancer diagnosis and treatment is of paramount importance. In 2019, over 228,000 new cases and over 140,000 estimated deaths are predicted in the US \cite{doi:10.3322/caac.21551}. The chance of surviving is higher when lung cancer is diagnosed in early cancer stages. The overall 5-year survival rate is approximately 70\% for people with stage \textit{IA/B} and 50\% for people with stage \textit{IIA/B} non-small lung cancer \cite{GOLDSTRAW2007706}. \\
In the past years, automated systems have been established to support the radiologists in diagnosing abnormalities on CXR images. Recent study shows that tremendous amount of nodule X-rays are required to compete with the nodule detection performance of radiologists \cite{CheXNet}. State-of-the art augmentation methods can be used to increase the amount of training data \cite{guan,CheXNet}. However, most of the augmentation methods hardly improve model performances as most techniques are applied on the whole image \cite{Shorten2019ASO}.\\

We present a method to extract nodules from the image and apply local, patch-based augmentation approaches to improve the system in nodule versus non-nodule image classification. A trained image inpainting network of CXR patches is used to replace a patch containing a nodule with authentic background structures. By subtracting the inpainted patch from the nodule patch, nodules can be separated from normal structures (e.g., tissues, bones, etc.). The extraction of nodules leads to various approaches which can be applied on the local nodule apart from the global CXR image. We show that a novel idea of augmentation - namely local feature augmentation - improves the system and can be defined as a better variant for nodule image augmentation based on CXRs.

\section{Background and Motivation}

\subsection{Computer-aided Systems on Pulmonary X-ray Scans}
In 2017, NIH released the first public CXR dataset with over 112,000 images and corresponding abnormality labels \cite{wang2017chestxray} which has led to various publications in classifying multiple abnormalities. The NIH group defines the baseline performance with an area under the curve (AUC) average of 0.75 across the abnormalities on their official evaluation set \cite{wang2017chestxray}. Hereafter, several groups increase the abnormality classification performance based on novel training strategies and network designs \cite{guan,DBLP:journals/corr/abs-1711-06373}. State-of-the-art results show that the performance for all 14 abnormalities raise to 0.82 AUC on average whereas the nodule classification score is improved to 0.78 AUC \cite{CheXNet}. However, over 6,000 nodule images are required to achieve such performance. Less training images significantly downgrade the performance as shown by Ausawalaithong et al. \cite{8609997}. Often, standard augmentation techniques are applied, e.g., horizontal flipping of the image \cite{guan,Yao2018LearningTD}. These methods imply some major drawbacks as the global modifications limit the degree of freedom to change the image. As nodules are small abnormalities with less than or equal to 30 millimeters in diameter, relevant features for nodule classification are only present locally, on a small fraction of the image. Accessing these local features of the image which contributes to the class prediction allows us to expand the augmentation space.

\subsection{Lung Region Inpainting for Classifier Deception}
Recent analysis shows that adversarial attacks can easily change classification predictions. Taghanaki et al. created a comprehensive evaluation how CXR abnormality classification networks act on adversarial perturbations \cite{CXR_Adv}. Given image data $D=\{x_1,x_2\dots x_n\}$ and its corresponding labels $Y=\{y_1,y_2\dots y_n\}$, a classification model $C$ can be trained by minimization of a loss function $l$. \eqref{eq-a}

\begin{subequations}
  \begin{tabularx}{11.5cm}{>{\hsize=5cm}Xp{1cm} X}

  \begin{equation}
    \label{eq-a}
      \argmin_C \sum_{x_i\in D}l(C(x_i),y_i)
  \end{equation}
  & &
  \begin{equation}
    \label{eq-b}
    \hspace{-0.1cm}
    \argmin_G\sum_{x_i\in D}l(C(G(x_i),y'_i))
  \end{equation}

  \end{tabularx}
\end{subequations}

The classification model can be attacked, e.g., with adversarial examples, generated from a model or method $G$ to change the prediction of model $C$ where $y_i\neq y'_i$. (\ref{eq-b}) 

Our novel approach builds on this basic idea generating realistic-looking patches in a supervised fashion which are placed in the image and change the nodule classification prediction. Local nodule features contributing to the class prediction which are covered by the inpainting frame can be entirely removed. The isolation process of the local features leads to various ideas; we focus on augmentation techniques to improve the classification system.

\section{Proposed Method}

\begin{figure}[h]
\centering
\includegraphics[width=12.2cm]{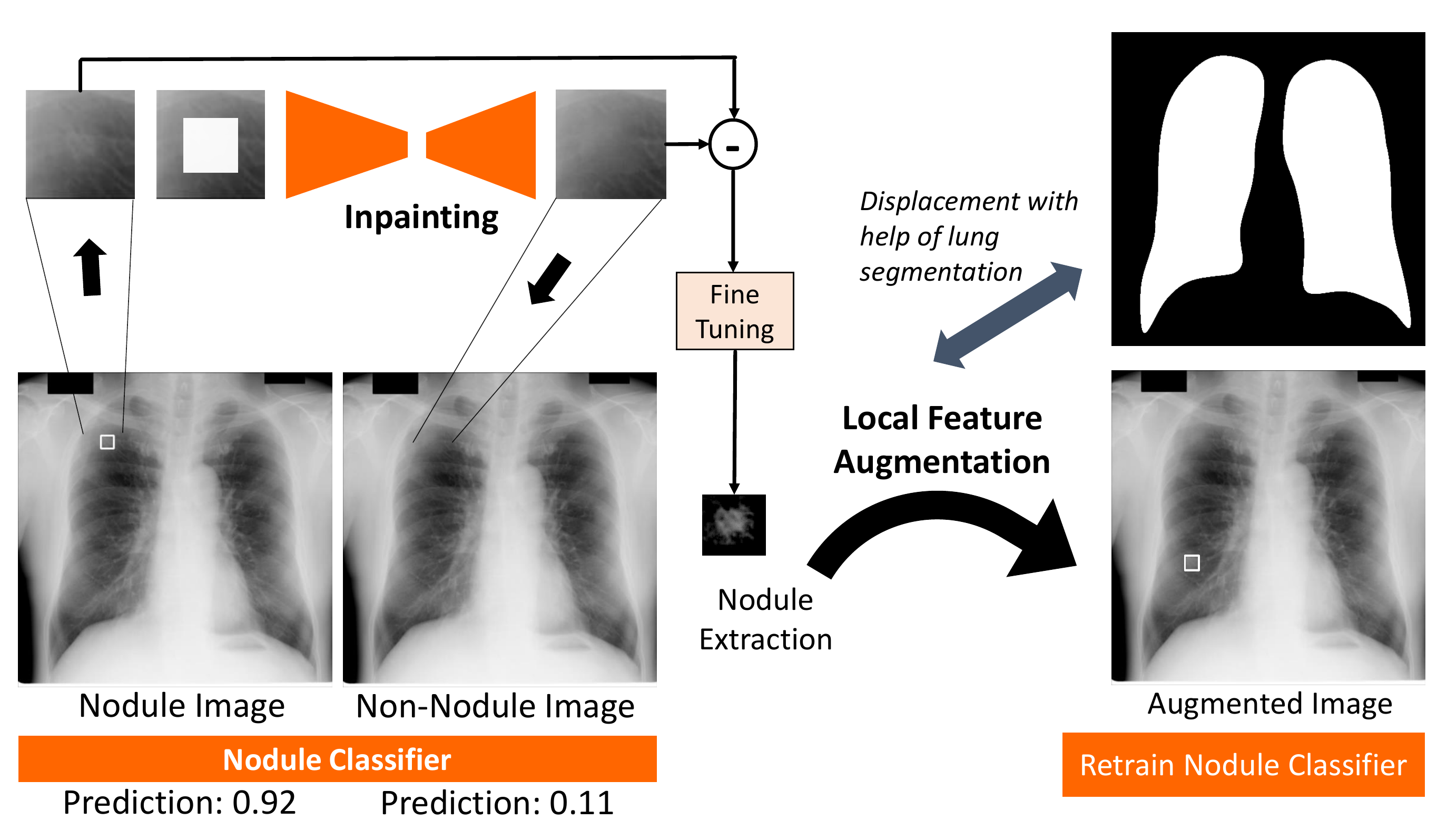}
\caption{Overview of our proposed method: We extract nodule patches from CXR images. The inpainting network predicts pixels for the mask. The nodule which has no correlation to the given frame surrounding that mask is entirely removed and lung tissue is predicted. Local augmentation techniques are applied on the isolated nodules to increase the classification performance. One augmentation method, the displacement of the nodule, can be arranged with given lung segmentation masks.}
\label{fig:overview}
\end{figure}

Our local feature augmentation system is composed of two parts. First, we extract nodule features using a patch inpainting method (left side of Figure \ref{fig:overview}). In a second step, with the help of the isolated nodules patches based on a nodule extraction process, we are able to displace nodules and apply further local augmentation techniques (right side). 

\subsection{Nodule Inpainting with Context Encoders}

Inspired by the work of Sogancioglu et al., we generate a patch inpainting network by using context encoders for our baseline model \cite{DBLP:journals/corr/abs-1809-01471}. 

We perform several modifications to improve the performance. As the context encoders do not perform well on predicting borders such that the rectangular contours of the mask are slightly visible after inpainting \cite{DBLP:journals/corr/abs-1809-01471}, we modify the system by applying the spatially discounted reconstruction loss introduced by Yu et al. \cite{Yu2018GenerativeII}. Missing pixels at the border have less ambiguity, hence, those pixels are weighted stronger during training. We predict the weight for each pixel with $\gamma^r$, where $r$ denotes the nearest distance to the mask border. As our mask size is smaller than the one in the reference work, we adapt $\gamma$ from 0.99 to 0.97. Further, we increase the network capacity of the network with channel size $c_l=2^{(8+l)}$ for encoder and adversarial part and $c_l=2^{(12-l)}$ for the decoder. The layer index $l$ is correspondingly $l_{enc}=\{0,1,2,3,4\}$ and $l_{dec/adv}=\{0,1,2,3\}$. From the full image, we extract patches with size $64\times 64$. A mask is overlayed with half as large in each dimension as the input patch.

After training, the nodules can be extracted with $n_i^{pre} = o_i - I(o_i^M)$ where $o_i$ denotes the original patches including the nodule and $o_i^M$ the same patches including the mask which is fed into the patch inpainting network $I$. Post-processing steps are applied to remove noises: $n_i = \Theta_s(max\{n_i^{pre},0\}), \forall i$. As pixels with nodules are brighter than pixels without nodules, we truncate all negative values to zero. Parameter $\Theta_s$ stands for the bilateral filter with a filter size of $s=3$. We hypothesize that the filter smooths the nodule patches and removes undesired background noise. 

\subsection{Local Feature Augmentation Techniques for Chest X-ray Classification}

The isolated nodule patches can be implanted in different ways to augment the dataset. In this study, we applied the extracted nodules on non-nodule images and not on the images where the nodules are derived from. In this way, the original nodule images are kept unchanged and more \textit{synthetic} nodule images are produced. According to this augmentation procedure, the dataset is more balanced with respect to the labels. 

Based on the given lung segmentation masks, we randomly insert a nodule patch in the lung region of an image with probability $k$. The corresponding class label is modified adequately. This process is performed for each epoch such that different images contain nodules during training. We hypothesize this variance enhancement procedure assists to make the classification model more robust during training. In addition to the nodule displacement, we apply further augmentation techniques to the local nodule patch: We use random rotation and flipping to achieve more variability. Because of the circular structure and the local properties of nodules we are able to rotate the patch by $r=[0,360[$ degrees, with $r\in \mathbb{N}_0$. The entire pipeline of the nodule extraction and augmentation process can be seen in pseudo code (Algorithm \ref{alg:pipeline}). \\

\begin{algorithm}[H]
 \KwData{nodule images $x$; bounding boxes $b$}
 \textbf{Networks:} Original Classification Network $C_1$; Retrained Classification Network $C_2$; Patch Inpainting Network $I$ \\
 \KwResult{Local Feature Augmentation for Classification Improvement}
 \For{$i\gets0$ \KwTo $len(x)$}{
    $o_{i}:=$ $getNodulePatch(x_{i},b_{i});$ \\
    $p_{i} := I(o_{i}^M)$ \tcp*{inpaint patch}
    $n_{i} := \Theta_s(max\{o_{i}-p_{i},0\});$\\
    \If{$C_1(patch2img(x_{i},p_{i})) < thr$}
    {
    consider $n_{i}$ for augmentation;\
       }
    }
 \While{Train($C_2$)}{
    \If{$random([0,1])<k$}{
    $AugmentImg(n_{rand})$ \tcp*{insert nodule patch to image}
    }
    train epoch with modified images;
 }
    
\caption{Local Feature Extraction for Classification Augmentation}
\label{alg:pipeline}
\end{algorithm}

\section{Experiments}

For our experiments, we use the ChestX-ray14 \cite{wang2017chestxray} and the JSRT \cite{JSRT} database. The combined database contains 112,367 images with 6,485 nodule images. Nodule bounding boxes for 233 images are provided in the datasets. We have lung segmentation masks available for all images retrieved from a standard U-Net segmentation network \cite{10.1007/978-3-319-24574-4_28}. The classification network is trained by using the architecture and hyperparameters from the work in \cite{10.1007/978-3-030-13469-3_88}. However, we upscale the input image size to $512\times 512$ to increase the resolution of the overall images and nodules.

\begin{figure}[h]
\centering
\includegraphics[width=12cm]{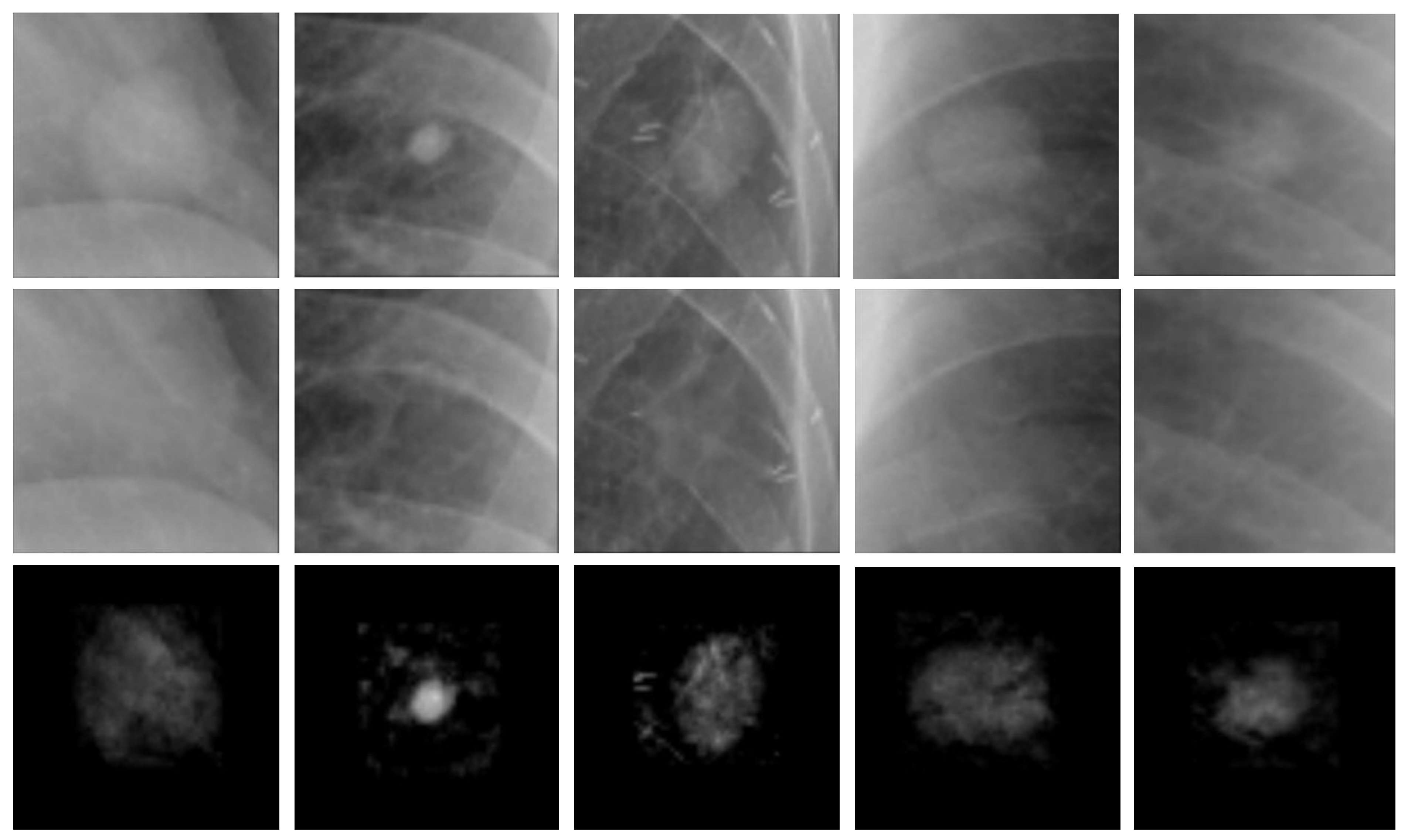}
\caption{Qualitative performance of our inpainting method for 4 different examples (Columns): Original nodule patch (Row 1); Inpainted Patch without nodule (Row 2); and subtracted patch including post-processing (Row 3)}
\label{fig:inpaint}
\end{figure}

At first, we trained the inpainting network $I$. We randomly collected 1 million, 10,000, and 800 patches for training, validation, and testing, respectively. The patches are extracted at random position from non-nodule images. The quantitative evaluation can be seen in Table \ref{tab:inp}.

\begin{table}[h]
\centering
\caption{Peak-Signal-to-Noise Ratio (PSNR) on a test set of 800 patches. CE: Context Encoder}
\label{tab:inp}
\begin{tabular}{C{2.5cm} | C{2.5cm} | C{2.5cm} | C{2.5cm}}
 &  CE \cite{DBLP:journals/corr/abs-1809-01471} &  CE* & CE* [ours] \\
\hline
PSNR \hspace{1cm} (mean $\pm$ std) & 26.31 $\pm$ 4.48 & 31.24$\pm$ 3.77 &\textbf{34.22 $\pm$ 3.95}\\
\multicolumn{4}{l}{*evaluated on a different test set}
\end{tabular}
\end{table}

In Figure \ref{fig:inpaint}, we illustrate 4 example patches (first row). The inpainted image can be seen in the second row and the subtracted patch after the post-processing steps in the third row. 

\begin{figure}[t]
\centering
\includegraphics[width=12.2cm]{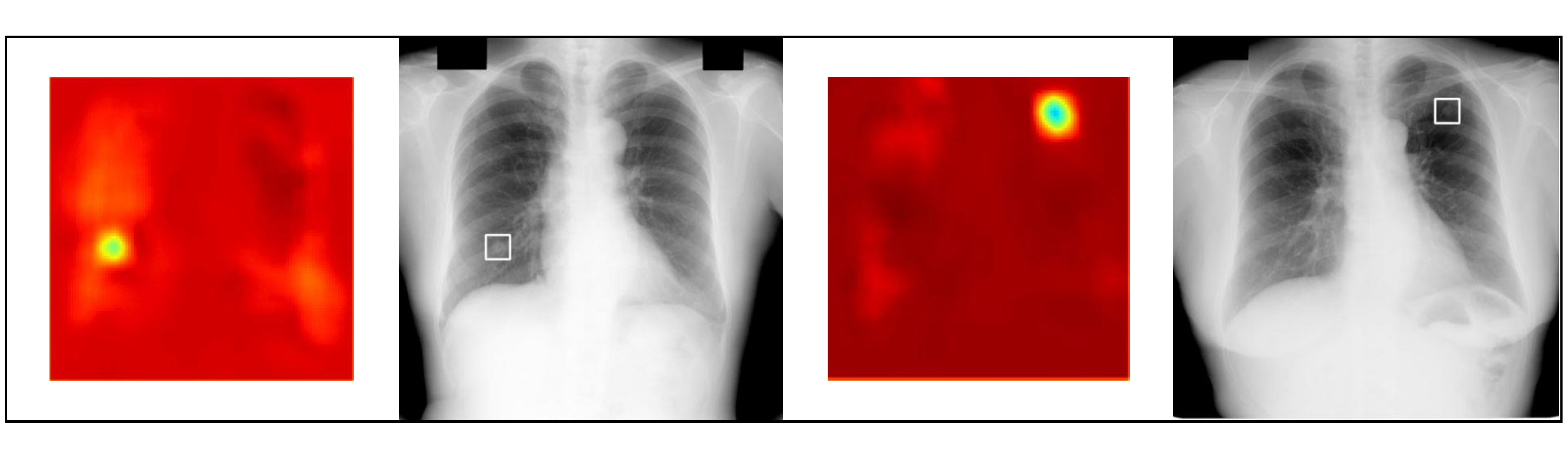}
\caption{Classification prediction visualized as attention: We sequentially inpaint all patches in nodule images and create attention maps based on the class prediction placed on the corresponding position.}
\label{fig:heat}
\end{figure}

In addition to the quantitative and qualitative evaluation in Table \ref{tab:inp} and Figure \ref{fig:inpaint}, we show attention maps in Figure \ref{fig:heat} based on 2 CXR images. Each possible patch of an image was inpainted sequentially, fed into the classification network, and the prediction is placed on the map. Figure \ref{fig:heat} shows that the image classification drops substantially when nodule regions were inpainted. We argue that the inpainting network is able to successfully remove nodules and replace it with background tissue to change the classifier prediction. Moreover, the prediction score remained stable when regions without nodules were inpainted, indicating that the inpainting network could generate normal patches robustly.

In order to ensure that nodules are reliably removed for augmentation purposes, the inpainting CXR images were individually validated. If the classification prediction was lower than the threshold $thr=0.5$, we considered the corresponding patch for the augmentation process. In addition to the training images, hence, we can include 178 nodule patches. The model was trained in following way: For each image and epoch we inserted a nodule patch with probability $k$. Accordingly, we changed the corresponding nodule label. 

To evaluate the benefit of using the local augmentation method on varying size of the training set, we performed learning curve analysis. We trained the network with t\% images of the training set and evaluated the performance. The dataset was split patient-wise into 70\%, 10\%, and 20\% for training, validation, and testing, respectively. We ensured that the images from the extracted nodule patches were present in the training set. For all experiments, we used a nodule insertion rate $k=0.05$. Each experiment was conducted 3 times. We show the resulting mean and standard deviation of the 3 runs in Table \ref{tab:class}. \\

\begin{table}[t]
\centering
\caption{AUC scores of the classification system. We show results based on different training set sizes for training the baseline model, adding standard augmentation and our proposed local features augmentation.}
\begin{tabular}{L{2.5cm} C{1.5cm} C{1.5cm} C{1.5cm} C{1.5cm} C{1.5cm} C{1.5cm}}
&\multicolumn{5}{c}{\textbf{Area Under the Curve}}\\
\cmidrule{2-7}
\multicolumn{1}{l}{Train set size [\%]}& 100 & 70 & 50 & 20 & 10 & 5\\
\multicolumn{1}{l}{Training images} & 79,011 & 55,307 & 39,505 & 15,802 & 7,901 & 3,950\\
\midrule
Baseline & 0.792\raisebox{.4ex}{\tiny{$\pm$0.010}} & 0.776\raisebox{.4ex}{{\tiny $\pm$0.012}} & 0.763\raisebox{.4ex}{{\tiny $\pm$0.009}} & 0.722\raisebox{.4ex}{{\tiny $\pm$0.019}} & 0.667\raisebox{.4ex}{{\tiny $\pm$0.007}} & 0.649\raisebox{.4ex}{{\tiny $\pm$0.009}}\\
Standard Augmentation &0.795\raisebox{.4ex}{\tiny{$\pm$0.004}} & 0.775\raisebox{.4ex}{{\tiny $\pm$0.008}} & 0.769\raisebox{.4ex}{{\tiny $\pm$0.010}} & 0.728\raisebox{.4ex}{{\tiny $\pm$0.013}} & 0.681\raisebox{.4ex}{{\tiny $\pm$0.005}} & 0.655\raisebox{.4ex}{{\tiny $\pm$0.007}}\\
\hdashline
\textbf{Local Feature Augmentation} &0.805\raisebox{.4ex}{\tiny{$\pm$0.004}} & 0.790\raisebox{.4ex}{{\tiny $\pm$0.005}} & 0.781\raisebox{.4ex}{{\tiny $\pm$0.004}} & 0.746\raisebox{.4ex}{{\tiny $\pm$0.005}} & 0.705\raisebox{.4ex}{{\tiny $\pm$0.017}} & 0.669\raisebox{.4ex}{{\tiny $\pm$0.013}}\\
\bottomrule
\end{tabular}
\label{tab:class}
\end{table}

We defined the baseline without any augmentation techniques (Row 1). Then, we conducted experiments with state-of-the-art augmentation on the full image. We applied random horizontal flipping and random rotation with a degree range of $[-15, 15]$. No significant improvement can be seen compared to the baseline model (Row 2). The evaluation of our local feature augmentation  method can be seen in Row 3. For each column the same training set was applied. For all training set sizes we state that our augmentation method consistently achieves better performance, compared to the baseline and standard augmentation method.

\section{Discussion}

In the proposed work, we demonstrated a novel image augmentation approach. We showed increased performance scores on nodule classification by applying augmentation locally. The advantage of the context model was the nearly unlimited training data as the local patches were retrieved from the full image. In this study, we collected 1 million patches for training, derived from a big data collection. However, only few images include information about nodule detection. Additionally, some nodules had to be sorted out during the retrieving process such that 178 nodules could be considered for the augmentation process. An additional radiologist validation process may decrease the amount of rejected samples as the current process was solely based on the classification system.\newline

The limited amount of nodules led to an improved nodule classification system on different training set sizes. In future work, more nodule annotations can be considered which may further increase the performance with local feature augmentation. The size of the full images was resized to 512 in each dimension. Especially small nodules may entirely disappear with this resolution change. Experiments on the original image size may result in an increased collection of valid nodule patches.\newline

We demonstrated the proposed method based on the classification of nodules. The system can be expanded to apply local feature augmentation on other local lung abnormalities, e.g., mass. Furthermore, we see no limitations to handle 3-dimensional image data, e.g., applying our local feature augmentation method on lung CT scans.

\section{Conclusion}

In this paper, we presented a novel idea of image augmentation to improve the nodule classification system on chest X-ray images. Instead of transforming the global image with standard augmentation techniques, we created an enhanced system of an inpainting model with context encoders to directly access nodule features. The extracted nodules were modified with our novel local feature augmentation method. Experiments conducted on different training set sizes showed significantly improved performance scores on nodule classification.

\ifx\anonymize\undefined
\textbf{Disclaimer} The concepts and information presented in this paper are based on research results that are not commercially available.
\fi

\bibliographystyle{splncs04}
\bibliography{main}

\end{document}